\documentclass{elsart}

\usepackage{graphicx}
\usepackage{amssymb}
\usepackage{amsmath}
\usepackage{amsfonts}
\usepackage{bbm} 
\usepackage{fleqn} 
\usepackage{indentfirst} 
\usepackage{braket}
\usepackage{mathrsfs}
\usepackage{HEP} 
\def\D{\mathrm{d}} 

\def\Tr{\text{Tr}}

\def\<{\left\langle}
\def\>{\right\rangle}

\begin{document}

\begin{frontmatter}

\begin{flushright}
{\small TUM-HEP-424/01}
\end{flushright}
\vspace*{2.cm}
\title{Neutrino Mass Operator Renormalization Revisited
}
\author{Stefan Antusch\thanksref{label01}},
\thanks[label01]{E-mail: \texttt{santusch@ph.tum.de}}
\author{Manuel Drees\thanksref{label02}},
\thanks[label02]{E-mail: \texttt{drees@ph.tum.de}}
\author{J\"{o}rn Kersten\thanksref{label04}
},
\thanks[label04]{E-mail: \texttt{jkersten@ph.tum.de}}
\author{Manfred Lindner\thanksref{label05}},
\thanks[label05]{E-mail: \texttt{lindner@ph.tum.de}}
\author{Michael Ratz\thanksref{label06}
}
\thanks[label06]{E-mail: \texttt{mratz@ph.tum.de}}
\address{Physik-Department T30, 
Technische Universit\"{a}t M\"{u}nchen\\ 
James-Franck-Stra{\ss}e,
85748 Garching, Germany
}

\begin{abstract}
We re-derive the renormalization group equation for the effective 
coupling of the dimension five operator which corresponds to a
Majorana mass matrix for the Standard Model neutrinos. We find
a result which differs somewhat from earlier calculations, 
leading to modifications in the evolution of leptonic mixing 
angles and CP phases.
We also present a general method for calculating $\beta$-functions 
from counterterms in MS-like renormalization schemes, which works 
for tensorial quantities.  
\end{abstract}

\end{frontmatter}

\newpage
\section{Introduction}
The Standard Model (SM) is most likely an effective theory 
up to some scale $\Lambda$, above which new physics 
has to be taken into account. The discovery of neutrino 
masses requires an extension of the SM, which 
may involve right-handed neutrinos or other new fields. 
Introducing right-handed neutrinos allows Dirac masses $m_\mathrm{D}$
via Yukawa couplings analogous to the quark sector. In general,
lepton number need not be conserved, so that Majorana masses
are possible. For left-handed neutrinos this can, 
for example, be achieved with Higgs triplets. Right-handed 
neutrinos can have explicit Majorana masses $M_\mathrm{R}$ 
of order $\Lambda$, 
since they are gauge singlets and since there are no protective 
symmetries. This leads to a picture with zero or tiny left-handed 
Majorana masses $M_\mathrm{L}$, with $m_\mathrm{D}$ similar to the 
charged lepton masses, and with a huge $M_\mathrm{R}$.
Diagonalization of the neutrino mass 
matrix results in Majorana fermions and eigenvalues 
$\simeq M_\mathrm{R}$ and 
$M_\mathrm{L} - m_\mathrm{D}^2/M_\mathrm{R}$. For $M_\mathrm{L}=0$ 
the neutrino masses are thus given by the see-saw 
relation $m_\mathrm{D}^2/M_\mathrm{R}$ \cite{seesaw}, which provides a 
convincing explanation for the smallness of neutrino masses.

Another, less model dependent approach 
is to study the effective field theory with
higher dimensional operators of SM fields. 
If lepton number is not conserved,
some of these generate Majorana neutrino 
masses. The lowest dimensional operator of this kind has 
dimension 5 and couples two lepton and two Higgs doublets. It
appears e.g. in the see-saw mechanism by integrating out the 
heavy right-handed neutrinos.

As quarks have only small mixings, it is somewhat surprising 
that neutrinos most likely have two large mixing angles 
\cite{Fogli:2001vr,Bahcall:2001zu,Bandyopadhyay:2001aa}. It is interesting to 
investigate mechanisms which can produce such large or maximal 
mixings. These mechanisms operate, however, typically at the 
embedding scale $\Lambda$. For a comparison of experimental 
results with high energy predictions from unified theories, it 
is thus essential to evolve the predictions to low energies 
with the relevant renormalization group equations (RGE's). 
This evolution is related to the running of the leading 
dimension 5 operator. Therefore, we calculate in this letter the 
RGE that governs this running above the electroweak scale
at one-loop order in the SM.

\section{Lagrangian and Counterterms}
Let $\ell_{\mathrm{L}}^f$, $f\in\{1,2,3\}$, be the 
SU(2)$_\mathrm{L}$-doublets of SM leptons,
$e_{\mathrm{R}}^f$ the SU(2)$_\mathrm{L}$-singlet
(right-handed) charged leptons, and
$\phi$ the Higgs doublet.
The dimension 5 operator that gives Majorana masses to
the SM neutrinos is given by
\begin{equation}\label{eq:Kappa:Babu:1993:1}
 \mathscr{L}_{\kappa} 
 =\frac{1}{4} 
 \kappa_{gf} \, \overline{\ell_\mathrm{L}^\mathrm{C}}^g_c\varepsilon^{cd} \phi_d\, 
 \, \ell_{\mathrm{L}b}^{f}\varepsilon^{ba}\phi_a  
  +\text{h.c.} \;,
\end{equation}
where $\kappa$ is symmetric under interchange of the generation indices
 $f$ and $g$, $\varepsilon$ is the totally antisymmetric tensor in 
2 dimensions, and 
$\ell_\mathrm{L}^\mathrm{C} := (\ell_\mathrm{L})^\mathrm{C}$ is 
the charge conjugate of the lepton doublet. $a,b,c,d \in \{1,2\}$ 
are SU(2) indices. They will only be written explicitly in terms 
with a non-trivial SU(2) structure. Summation over repeated indices 
is implied throughout this letter.

\begin{figure}[h]
\[
        \vcenter{\hbox{\includegraphics{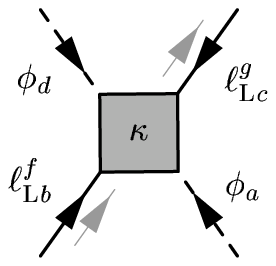}}}
        \quad : \quad
        i \kappa_{gf} \, \tfrac{1}{2} ( 
         \varepsilon^{cd} \varepsilon^{ba} + 
         \varepsilon^{ca} \varepsilon^{bd} ) 
\]
\caption{Vertex from the effective dimension 5 operator and the 
corresponding Feynman rule. The gray arrow indicates the fermion flow 
as defined in \cite{Denner:1992vz}.} \label{fig:KappaVertex}
\end{figure}

$\mathscr{L}_{\kappa}$ gives rise to the vertex shown in 
Fig.~\ref{fig:KappaVertex}, and an analogous one for the Hermitian 
conjugate term.

The complete Lagrangian consists of $\mathscr{L}_{\kappa}$, the 
SM Lagrangian $\mathscr{L}_\mathrm{SM}$ and proper 
counterterms $\mathscr{C}$,
\begin{equation}
        \mathscr{L} =
         \mathscr{L}_\kappa +\mathscr{L}_\mathrm{SM} + \mathscr{C} \;.
\end{equation}
In the following, we omit most of those parts that yield only flavour 
diagonal contributions to the $\beta$-function and therefore do not 
contribute to the running of mixing angles, in particular terms 
involving quarks and gauge bosons. The remaining ones are
\begin{subequations}
\begin{eqnarray}
        \mathscr{L}_{\mathrm{kin}(\ell_\mathrm{L})} &=&
         \overline{\ell_\mathrm{L}}^f (i\gamma^\mu \partial_\mu) 
         \ell_\mathrm{L}^f \;,
        \\
        \mathscr{L}_\mathrm{Higgs} &=&
         (\partial_\mu \phi)^\dagger (\partial^\mu \phi)
                 - m^2 \, \phi^\dagger \phi
         - \tfrac{1}{4} \lambda \, (\phi^\dagger \phi)^2 \;,
        \\
        \mathscr{L}_\mathrm{Yukawa} &=&
         - (Y_e)_{gf} \, \overline{e_\mathrm{R}}^g \phi^\dagger 
         \ell_\mathrm{L}^f + \text{h.c.} \;;
\end{eqnarray}
\end{subequations}
\begin{subequations}
\begin{eqnarray}
        \mathscr{C}_{\mathrm{kin}(\ell_\mathrm{L})} &=&
         \overline{\ell_\mathrm{L}}^g (i\gamma^\mu \partial_\mu)
         (\delta Z_{\ell_\mathrm{L}})_{gf} \ell_\mathrm{L}^f \;,
        \\
        \mathscr{C}_\mathrm{Higgs} &=&
         \delta Z_\phi (\partial_\mu \phi)^\dagger(\partial^\mu \phi)
                  - \delta m^2 \, \phi^\dagger \phi
         - \tfrac{1}{4} \, \delta\lambda (\phi^\dagger \phi)^2\;,
        \\
                \mathscr{C}_\mathrm{Yukawa} &=&
                - (\delta Y_e)_{gf} \, \overline{e_\mathrm{R}}^g \phi^\dagger 
         \ell_\mathrm{L}^f + \text{h.c.} \;;
\end{eqnarray}
\end{subequations}
\begin{equation}
        \mathscr{C}_\kappa =
         \tfrac{1}{4} \delta \kappa_{gf} \,
         \overline{\ell_\mathrm{L}^\mathrm{C}}^g_c \varepsilon^{cd} \phi_d \,\, 
         \ell_{\mathrm{L}b}^f \varepsilon^{ba} \phi_a + \text{h.c.} \;.
\end{equation}
$\delta Z_i$ ($i \in \{\ell_\mathrm{L}, \phi\}$) determine the 
wavefunction renormalization constants 
$Z_i = \mathbbm{1} + \delta Z_i$, defined in the usual way. Note that 
$Z_{\ell_\mathrm{L}}$ is a matrix in flavour space.
$\delta \kappa$ satisfies the relation
\begin{equation}
        \kappa_\mathrm{B} = 
        Z_\phi^{-\frac{1}{2}}
        \left( Z_{\ell_\mathrm{L}}^T \right)^{-\frac{1}{2}} \,
        \left[ \kappa+\delta \kappa \right] \, \mu^\epsilon\,
        Z_{\ell_\mathrm{L}}^{-\frac{1}{2}}
        Z_\phi^{-\frac{1}{2}} \;,
\end{equation}
where the factor $\mu^\epsilon$ is due to dimensional
regularization, with $\mu$ denoting the renormalization scale and 
$\epsilon := 4-d$. The subscript B denotes a bare quantity. 
Note that the usual ansatz $\kappa_\mathrm{B} \sim Z_\kappa \kappa$ 
is not possible in this case, as it would obviously spoil the symmetry 
of $\kappa_\mathrm{B}$ or $\kappa$ with respect to interchange of the 
flavour indices.

\section{Calculation of the Counterterms}

\begin{figure}[h]
\begin{eqnarray*}
\lefteqn{
\vcenter{\hbox{\includegraphics{Letter1_13.eps}}}
+
\vcenter{\hbox{\includegraphics{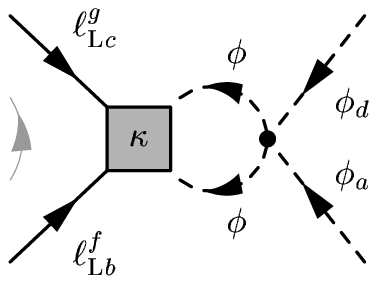}}}
+
\vcenter{\hbox{\includegraphics{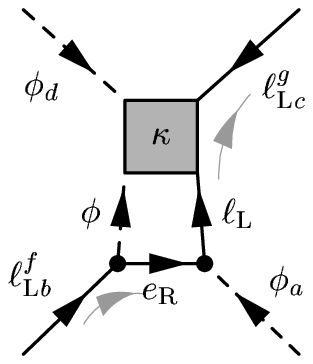}}}
}
\\
&&
+
\vcenter{\hbox{\includegraphics{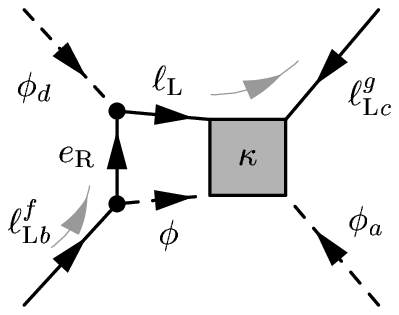}}}
+
\vcenter{\hbox{\includegraphics{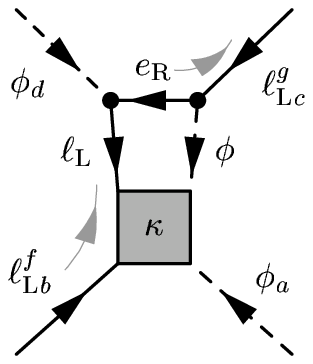}}}
+
\vcenter{\hbox{\includegraphics{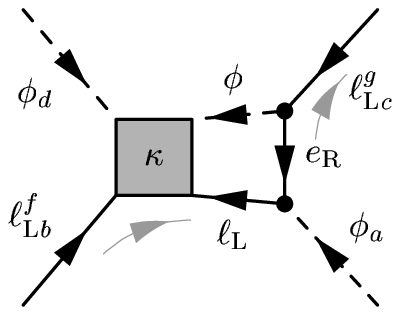}}}
\\
&&
+
\vcenter{\hbox{\includegraphics{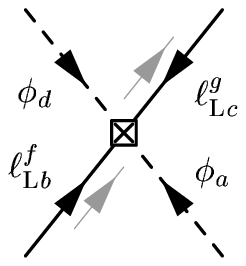}}}
\quad+\quad \mbox{diagrams from gauge interactions}
\end{eqnarray*}
\caption{\label{fig:DiagramsForKappaRen} Diagrams relevant for the 
renormalization of the vertex from the effective dimension 5 neutrino 
mass operator. The last diagram represents the counterterm.}
\end{figure}

In the MS scheme, the quantity $\delta \kappa$ can be computed at 
one-loop order from the requirement that the sum of diagrams in 
Fig.~\ref{fig:DiagramsForKappaRen} be ultraviolet finite.

Using FeynCalc \cite{Mertig:1991an} we obtain 
\begin{equation}\label{eq:deltazkappa}
        \delta \kappa = -\frac{1}{16\pi^2} \left[
         2(Y_e^\dagger Y_e)^T \kappa + 2\kappa\,(Y_e^\dagger Y_e)
         -\lambda \kappa + C_\kappa 
        \right] \frac{1}{\epsilon} \;,
\end{equation}
where $C_\kappa$ denotes the contribution from gauge interactions.
The usual calculation of the wavefunction renormalization constants 
yields
\begin{eqnarray}
        \delta Z_\phi & = & -\frac{1}{8\pi^2} \left[
         \Tr (Y_e^\dagger Y_e) + C_\phi 
        \right] \frac{1}{\epsilon} \;, \label{eq:deltaZphi}
        \\
        \delta Z_{\ell_\mathrm{L}} & = & -\frac{1}{16\pi^2} \left[
         Y_e^\dagger Y_e + C_{\ell_\mathrm{L}} \mathbbm{1}
        \right] \frac{1}{\epsilon} \;.\label{eq:deltaZlL}
\end{eqnarray}
Again, $C_\phi$ and $C_{\ell_\mathrm{L}}$ represent terms 
from quarks and gauge interactions, which are diagonal in flavour space.

\section{Calculating RGE's from Counterterms with Tensorial Structure}

The calculation of the \(\beta\)-function involves some subtle points,
which are related to the matrix structure of the counterterm Lagrangian.
Before presenting our result in
Sec.~\ref{sec:RGE}, we provide now some details of the calculation, which should be of general
interest and which are essential for verifying our result.
In particular, we generalize the usual formalism for calculating
$\beta$-functions to include tensorial quantities as well as
non-multiplicative renormalization. 

We are interested in the \(\beta\)-function for a quantity \(Q\),
$ \beta_Q := \mu \frac{\D Q}{\D\mu}$.
In general, the bare and the renormalized quantity are related by
\begin{eqnarray}
 Q_\mathrm{B} 
 & = &
 Z_{\phi_1}^{n_1}\cdots Z_{\phi_M}^{n_M}\,
 [Q+\delta Q]\mu^{D_Q\epsilon}
 \,Z_{\phi_{M+1}}^{n_{M+1}}\cdots Z_{\phi_N}^{n_N}
 \nonumber\\
 & = &
 \left(\prod_{i\in I}Z_{\phi_i}^{n_i}\right)\,
 [Q+\delta Q]\mu^{D_Q\epsilon}
 \,\left(\prod_{j\in J}Z_{\phi_j}^{n_j}\right)\;,
 \label{eq:AddRenQB}
\end{eqnarray}
where \(I=\{1,\dots, M\}\), \(J=\{M+1,\dots, N\}\) and $D_Q$ is related to the
mass dimension of $Q$. 
\(\delta Q\) and the wavefunction renormalization constants depend on \(Q\) and some
additional variables \(\{V_A\}\),
\begin{subequations}
\begin{eqnarray}
 \delta Q & = & \delta Q(Q,\{ V_A\})\;,\\
 Z_{\phi_i} & = & Z_{\phi_i}(Q,\{V_A\})
 \qquad (1\le i\le N)\;.
\end{eqnarray}
\end{subequations}
Note that \(Q = Q (\mu)\) and \(V_A = V_A (\mu)\) are functions of the renormalization 
scale $\mu$, but 
 \(\delta Q\) and \(Z_{\phi_i}\) do not depend explicitly on $\mu$ in an MS-like
 renormalization scheme.
Taking the derivative of equation (\ref{eq:AddRenQB}) yields
\begin{eqnarray}\label{eq:BetaFtnDer}
 0 & \stackrel{!}{=} &
 \mu^{-D_Q\epsilon}\mu\frac{\D}{\D \mu} Q_\mathrm{B}
 \nonumber\\
 & = & 
 \left(\prod_{i\in I}Z_{\phi_i}^{n_i}\right)\,
 \left[\beta_Q 
        +\Braket{\frac{\D\delta Q}{\D Q}|\beta_Q}+\right.
  \nonumber\\
  & & 
  \left.
       \qquad +\sum_A\Braket{\frac{\D\delta Q}{\D V_A}|\beta_{V_A}}
        +\epsilon D_Q(Q+\delta Q)
        \right]
 \,\left(\prod_{j\in J}Z_{\phi_j}^{n_j}\right)
 \nonumber\\
 & & 
 + \left(\prod_{i\in I}Z_{\phi_i}^{n_i}\right)\,
 [Q+\delta Q]\left\{\sum_{j\in J}
        \left(\prod_{j'<j}Z_{\phi_{j'}}^{n_{j'}}\right)\right.\times
 \nonumber
 \\
 &&\qquad
 \left.\times\,
        \left[\Braket{\frac{\D Z_{\phi_j}^{n_j}}{\D Q}|\beta_Q}
                +\sum_A\Braket{\frac{\D Z_{\phi_j}^{n_j}}{\D V_A}|\beta_{V_A}}\right]
        \left(\prod_{j''>j}Z_{\phi_{j''}}^{n_{j''}}\right)\right\}
 \nonumber\\
  & &
 +
 \left\{\sum_{i\in I}
        \left(\prod_{i'<i}Z_{\phi_{i'}}^{n_{i'}}\right)
        \left[\Braket{\frac{\D Z_{\phi_i}^{n_i}}{\D Q}|\beta_Q}
                +\sum_A\Braket{\frac{\D Z_{\phi_i}^{n_i}}{\D V_A}|\beta_{V_A}}\right]
                \right.
        \times
\nonumber\\
 & &
 \qquad\times
 \left.\left(\prod_{i''>i}Z_{\phi_{i''}}^{n_{i''}}\right)\right\}\,
 [Q+\delta Q]
 \,\left(\prod_{j\in J}Z_{\phi_j}^{n_j}\right)
 \;.
\end{eqnarray}
Here we have introduced the notation
\begin{equation}
\Braket{\frac{\D F}{\D x}|y} := 
 \left\{
 \begin{array}{ll}
        \displaystyle\frac{\D F}{\D x}y
        &\text{for scalars}\;x,y\\
        \displaystyle\sum_{n}\frac{\D F}{\D x_{n}} y_{n}
        \quad
        &\text{for vectors}\;x=(x_{m}),y=(y_{m})\\
        \displaystyle\sum_{m,n}\frac{\D F}{\D x_{mn}} y_{mn}
        \quad
        &\text{for matrices}\;x=(x_{mn}),y=(y_{mn})\\
        \dots
        &\text{etc.}\;.
 \end{array}\right.
\end{equation}
We will solve equation (\ref{eq:BetaFtnDer}) and the corresponding expression
for $V_A$ by expanding all quantities in powers of $\epsilon$. In the
MS-scheme the quantities $\delta Q$ and $Z_{\phi_i}$ can be expanded as
\begin{subequations}\label{eq:AddRenAssumptions1}
\begin{eqnarray}
 \delta Q 
 & = & \sum_{k\ge 1} \frac{\delta Q_{,k}}{\epsilon^k}
 \;,\\
 Z_{\phi_i}
 & = & \mathbbm{1}+\sum_{k\ge 1}\frac{\delta Z_{\phi_i,k}}{\epsilon^k}
 =: \mathbbm{1}+\delta Z_{\phi_i} \;,
\end{eqnarray}
\end{subequations}
with higher powers of $\frac{1}{\epsilon}$ corresponding to higher powers in
perturbation theory. On the other hand, $\beta$-functions are finite as
$\epsilon \rightarrow 0$. We can therefore make the ansatz
\begin{subequations}\label{eq:AddRenAssumptions2}
\begin{eqnarray}
\beta_Q 
 & = & \beta_Q^{(0)} + \epsilon \beta_Q^{(1)} + \dots + \epsilon^n\beta_Q^{(n)}
 \;,\\
 \beta_{V_A} 
 & = & \beta_{V_A}^{(0)} + 
        \epsilon \beta_{V_A}^{(1)} + \dots + \epsilon^n\beta_{V_A}^{(n)}
 \;,
\end{eqnarray}
\end{subequations}
where $n$ is an arbitrary integer. Note that in this case the power of
$\epsilon$ is not related to the order of perturbation theory.
From (\ref{eq:AddRenAssumptions1}) and (\ref{eq:AddRenAssumptions2}) we find
that
\begin{equation}\label{eq:DZ}
 \frac{\D Z_{\phi_i}^{n_i}}{\D Q} 
 = n_i\, Z_{\phi_i}^{n_i-1}\frac{\D Z_{\phi_i}}{\D Q}
 = n_i \frac{\D \delta Z_{\phi_i}}{\D Q}
        +\mathscr{O}\left(\tfrac{1}{\epsilon^2}\right)
 =\mathscr{O}\left(\tfrac{1}{\epsilon}\right)\; ,
\end{equation}
where the lowest possible power of $\frac{1}{\epsilon}$ appearing on the 
right side of (\ref{eq:DZ}) is $1$.
An analogous relation holds for \(Q\leftrightarrow V_A\). 
Our analysis of equation (\ref{eq:BetaFtnDer}), starting with the 
inspection of the $\epsilon^n$ term, then shows that
$\beta_Q^{(n)}$ vanishes. The analog of equation (\ref{eq:BetaFtnDer}) for
$\beta_{V_A}$ implies that $\beta_{V_A}^{(n)}$ vanishes as well. Repeating this
argument for successively smaller positive powers of $\epsilon$ implies that
\begin{subequations}
\begin{eqnarray}
 \beta_Q^{(k)} & = &\beta_{V_A}^{(k)}=0\qquad \forall \;k\in \{2,\dots,n\}\;, \\
 \beta_Q^{(1)} & = & - \epsilon D_Q Q \;,\\
 \beta_{V_A}^{(1)}  & = & - \epsilon D_{V_A} V_A \;.
\end{eqnarray} 
\end{subequations}
Note that these terms do not contribute to the $\beta$-function in 4 dimensions,
i.e. for $\epsilon \rightarrow 0$, but they are necessary to read off 
$\beta_Q^{(0)}$ 
from equation (\ref{eq:BetaFtnDer}), leading to the result
\begin{eqnarray}\label{formula}
 \beta^{(0)}_Q & = & 
 \left[
        D_Q\Braket{\frac{\D\delta Q_{,1}}{\D Q}|Q}
        +\sum_A  D_{V_A}\Braket{\frac{\D\delta Q_{,1}}{\D V_A}|V_A}
        -D_Q\,\delta Q_{,1}
         \right]
 \nonumber\\
 & &
 +Q\cdot\sum_{j\in J} n_j
        \left[D_Q\,\Braket{\frac{\D Z_{\phi_j,1}}{\D Q}|Q}
                +\sum_A D_{V_A}\,\Braket{\frac{\D Z_{\phi_j,1}}{\D V_A}|V_A}\right]
 \nonumber\\
  & &
 +\sum_{i\in I}n_i
        \left[D_Q\,\Braket{\frac{\D Z_{\phi_i,1}}{\D Q}|Q}
                +\sum_A D_{V_A}\,\Braket{\frac{\D Z_{\phi_i,1}}{\D V_A}|V_A}\right]
 \cdot Q
 \;.
 \label{eq:BetaFunctionInAdditionalRenormalization}
\end{eqnarray}
Note that
for complex quantities \(Q\) and \(V_A\) we have to treat the complex conjugates
\(Q^*\) and \(V_A^*\) as additional independent variables.

\section{Renormalization Group Equation} \label{sec:RGE}
The RGE for the effective coupling $\kappa$ is
\begin{eqnarray}
        \mu \frac{\D \kappa}{\D \mu} = \beta_\kappa \;.
\end{eqnarray}
Using equations (\ref{formula}) and
(\ref{eq:deltazkappa}) -- (\ref{eq:deltaZlL}),
 we obtain for the contributions from
vertex and wavefunction renormalization 
(omitting terms from $C_\kappa, C_\phi$ and $C_{\ell_\mathrm{L}}$):
\begin{subequations}
\begin{eqnarray} \label{eq:rgeparts}
        16\pi^2 \beta_\kappa^{\text{(v)}} &=&  
         -2 \left[
                  \kappa  (Y_e^\dagger Y_e) + (Y_e^\dagger Y_e)^T \kappa 
                 \right]
         +\lambda \kappa \;,
        \\
        16\pi^2 \beta_\kappa^{\text{(wf)}} &=&  
         \frac{1}{2} \left[\kappa \, (Y_e^\dagger Y_e) +
                       (Y_e^\dagger Y_e)^T \kappa \right]+
         2 \, \Tr (Y_e^\dagger Y_e) \kappa \;.
\end{eqnarray}
\end{subequations}
Adding the terms involving quarks and gauge bosons 
\cite{Babu:1993qv,Chankowski:1993tx}, we obtain the final result
\begin{eqnarray} \label{eq:finalrge}
        16\pi^2 \beta_\kappa & = &
         -\frac{3}{2} \left[\kappa \left( Y_e^\dagger Y_e \right)
         +            \left( Y_e^\dagger Y_e \right)^T \kappa \right]
         +
        \nonumber\\
         & &
         +\lambda \kappa - 3 g_2^2 \kappa
         +2 \, \Tr \left( 3 Y_u^\dagger Y_u + 3 Y_d^\dagger Y_d 
         +Y_e^\dagger Y_e \right) \kappa \;,
\end{eqnarray}
where $g_2$ is the SU(2) gauge coupling constant and where $Y_u$, 
$Y_d$ are the Yukawa matrices for the up and the down quarks.
Thus, compared to earlier results \cite{Babu:1993qv}, we find a coefficient
$-\frac{3}{2}$ instead of $-\frac{1}{2}$ in front of the non-diagonal
term $\kappa (Y_e^\dagger Y_e) + (Y_e^\dagger Y_e)^T \kappa$.
Note that the difference in the $\lambda\kappa$-term is due to a 
different convention for the Higgs self-interaction used in this work.

We have checked our results by calculating the essential parts of
the same $\beta$-functions 
from the finite parts of the relevant diagrams 
in the framework of an underlying
renormalizable theory. 
This calculation as well as the application to  
the MSSM and the two Higgs SM will be presented in a 
future paper \cite{ADKL:2001b}.

\section{Discussion and Conclusions}
We have calculated in the SM the $\beta$-function for the effective 
coupling $\kappa$ of the dimension 5 operator which corresponds to a 
Majorana mass matrix for neutrinos. 
We have explicitly 
presented our calculations for the non-diagonal part of the 
$\beta$-function, where our result disagrees with the
previous one in \cite{Babu:1993qv} by a factor of 3. 
This part is responsible for the evolution of 
neutrino mixing angles and CP phases.
Therefore, our result modifies 
the renormalization group running of these quantities between 
predictions of models at high energies and experimental 
data at low energies. 
Consequently, our work affects the SM results of previous studies 
based on the existing RGE's, e.g.\ 
\cite{Ellis:1999my,Ibarra:1999se,Casas:1999tg,Balaji:2000gd,Balaji:2000ma,Kuo:2001ha}.

\ack
We would like to thank A.~Buras, W.~Grimus and Ch.~Wetterich 
for useful discussions. This work was supported by the 
``Sonderforschungsbereich~375 f\"ur Astro-Teilchenphysik der 
Deutschen Forschungsgemeinschaft''.
M.R. acknowledges support from the ``Promotionsstipendium des Freistaats Bayern''.

\end{document}